\documentclass[twoside,leqno,twocolumn]{article}
\usepackage{ltexpprt}

\usepackage{amsmath,bm,graphicx,amsfonts}
\usepackage{caption}
\usepackage{subcaption}
\usepackage{lipsum}

\makeatletter
\def\hlinewd#1{%
  \noalign{\ifnum0=`}\fi\hrule \@height #1 \futurelet
   \reserved@a\@xhline}
\makeatother

\begin{document}
\title{\Large Making \#Sense of \#Unstructured Text Data$^*$
\author{
Lin Li, William M. Campbell, Cagri Dagli, Joseph P. Campbell \\
MIT Lincoln Laboratory\\
\texttt{lin.li@ll.mit.edu, wcampbell@ll.mit.edu}
}}
\date{}
\maketitle

\renewcommand{\thefootnote}{*} \footnotetext{This work was sponsored
  by the Defense Advanced Research Projects Agency under Air Force
  Contract FA8721-05-C-0002. Opinions, interpretations, conclusions,
  and recommendations are those of the authors and are not necessarily
  endorsed by the United States Government.}
\maketitle

\begin{abstract}
Many network analysis tasks in social sciences rely on pre-existing
data sources that were created with explicit relations or interactions
between entities under consideration. Examples include email logs,
friends and followers networks on social media, communication
networks, etc. In these data, it is relatively easy to identify who is
connected to whom and how they are connected. However, most of the
data that we encounter on a daily basis are unstructured free-text
data, e.g., forums, online marketplaces, etc. It is considerably more
difficult to extract network data from unstructured text. In this
work, we present an end-to-end system for analyzing unstructured text
data and transforming the data into structured graphs that are
directly applicable to a downstream application. Specifically, we
look at social media data and attempt to predict the most indicative
words from users' posts. The resulting keywords can be used to
construct a context+content network for downstream processing such as
graph-based analysis and learning. With that goal in mind, we apply
our methods to the application of cross-domain entity resolution. The
performance of the resulting system with automatic keywords shows
improvement over the system with user-annotated hashtags.
\end{abstract}

\section{Introduction}

Automatic extraction of intelligent and useful information from data
is one of the main goals in data science. Traditional approaches have
focused on learning from structured features, i.e., information in a
relational database. However, most of the data encountered in practice
are unstructured (i.e., forums, emails and web
logs); they do not have a predefined schema or format. The challenge
of learning and extracting relevant pieces of information from
unstructured text data is an important problem in natural language
processing and can render useful graph representation for various
machine learning tasks.

Various efforts have been proposed to develop algorithms for processing
unstructured text data.  At a top level, text can be either summarized
by document level features (i.e., language, topic, genre, etc.) or
analyzed at a word or sub-word level.  Text analytics can be
unsupervised, semi-supervised, or supervised. 
 
In this work, we focus on word analysis and examine {\it unsupervised}
methods for processing unstructured text data, extracting relevant
information, and transforming it into structured relations that
can then be leveraged in graph analysis and various other downstream
applications. Unsupervised methods require less
human annotation and can easily fulfill the role of automatic
analysis. The specific application that we examine in this work is the
problem of associating entities (typically people and organization)
across multiple platforms, also known as the {\it cross-domain entity
resolution}~\cite{campbell2016cross}.

Consider social media platforms as an example. Information about an
entity is often present in multiple forms, such as profile, social
activities and content. Profiles give information about the username,
full name, profile pictures, etc. Social activities (i.e., mentions,
comments, re-posts, etc.) provide information about the interaction
among entities on social media. Content covers almost everything a
user posts on social media. One prominent example is the use of
hashtags. Hashtags can be considered as user-annotated keywords that
are indicative of the content or the topic of the posts.  Therefore,
hashtags are often introduced in the graph construction process that
mix together content features and context features. The resulting
context+content graph is often able to provide better and more
interpretable results. Yet hashtags are generally noisy because there
is no restriction on how they can be used in a
post~\cite{davidov2010enhanced}. For example, users may mark the word
{\tt boston} as a hashtag (i.e., {\tt \#boston}) in one post, but not
use it as a hashtag in other posts. Additionally, some users tend to
use too many hashtags in a post while others may not use hashtags at
all. In the case of forums, emails, web logs or online marketplaces,
there simply exists no user-annotated hashtags.

The goal of this work is to investigate methods for learning a graph
representation of noisy and unstructured text data via automatic
hashtag annotation and contrast it to standard methods which are based
on user-annotated hashtags. Automatic hashtag annotation is a way of
consistently extracting indicative content words from the posts. These
words are then used to construct a graph relation based on user
mentions, co-occurrences and user interactions. We refer to the
resulting graph as a content+context graph. Automatic hashtag
annotation is closely related to automatic tag
extraction~\cite{krestel2009latent,liu2011simple} and keyword
extraction~\cite{mihalcea2004textrank,zhao2011topical}. Although here
we focus on the problem of structuring social media data, the methods
for automatic hashtag annotation and representing the data in graph
form are more general and also applicable to non-social media
platforms, such as forums, emails, and marketplaces.

\section{Structuring Unstructured Text Data}

\begin{figure}[h]
\centering
\includegraphics[width=0.4\textwidth]{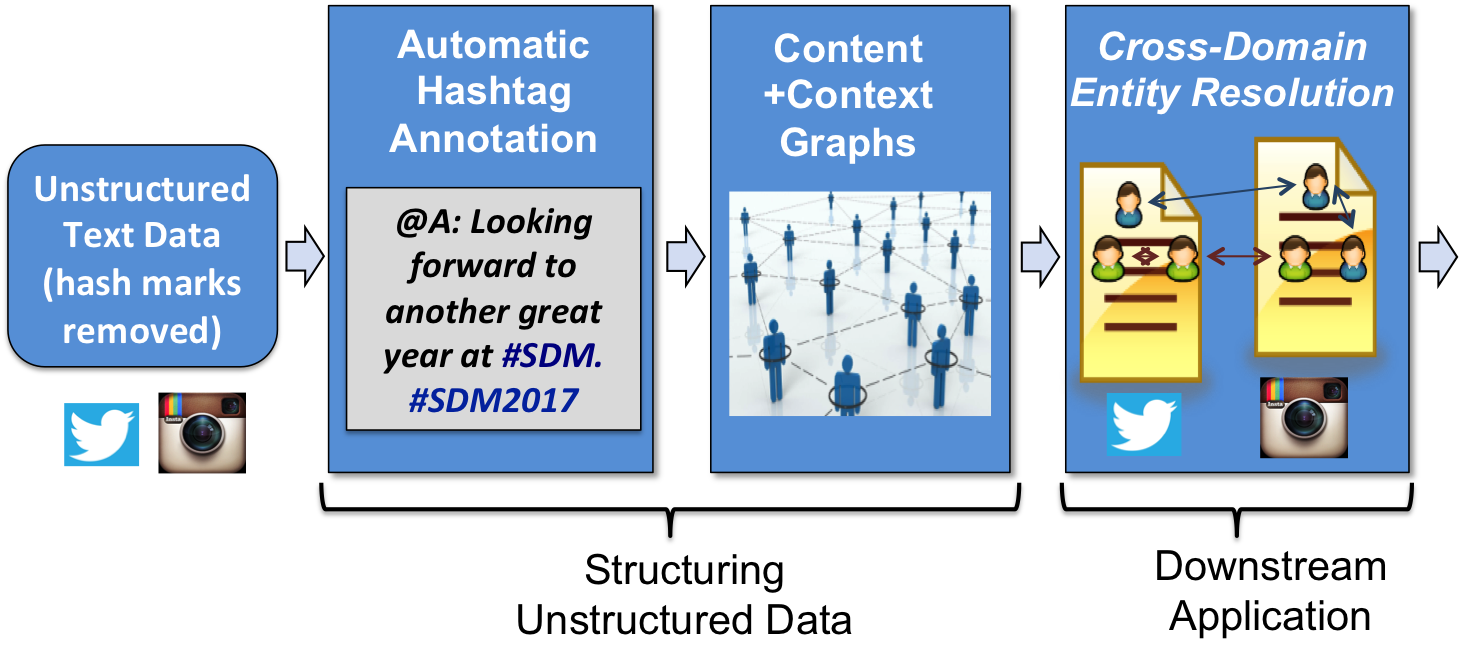}
\caption{Proposed framework for unstructured text analysis}
\label{fig:system}
\vspace{-0.2in}
\end{figure}

The proposed framework for structuring unstructured text data is shown
in Figure~\ref{fig:system}. It differs from previous methods by
offering an end-to-end process of analyzing unstructured text data via
automatic hashtag annotation, building graphs of entities and
hashtags, and performing graph analyses that are directly applicable
to downstream applications. For text analysis, we focus on methods for
finding relevant words in the text.  Specifically, we look at social
media data (Twitter and Instagram) and attempt to predict indicative
content words for users' posts.  The resulting words are then used for
constructing graphs that are relevant to the downstream
application. The specific application that we examine in this work is
the problem of associating entities (typically people and
organization) across multiple social media sites, known as
cross-domain entity resolution; see Section \ref{sec:ERS}.

\subsection{Data Sets}\label{section:data}
Twitter and Instagram data from the Boston area were
collected for our experiments~\cite{Li2015,campbell2016cross}. Twitter
data consist of approximately $4.65$ million tweets collected from
$1/1/2014$ to $10/1/2014$. Instagram data consist of $3.71$ million
posts (and comments) collected between $12/31/2013$ to
$12/31/2014$. For Instagram, some comments on these posts extend into
$2015$.

\subsection{Pre-processing}\label{sec:preproc}
Our pipeline for pre-processing the social media content data is as
follows. First, we perform text normalization on the string; it
converts any unusual UTF-8 characters to a standardized form,
eliminates emojis, emoticons, links and retweets. Then we apply
standard text pre-processing steps: tokenization, punctuation and stop
word removal. Lastly, we remove all the hash marks from the string. Hence, what used to be a hashtag is now a regular word in the post\footnote{Although a hashtag may contain multiple words, in this work, we treat it as a phrase with no space separating the words.}. 

\subsection{Automatic Hashtag Annotation}
For hashtag annotations, we explore two different but related
strategies: a topic-based method and a community-based graph method. 

Our pipeline for the topic-based method is as follows. First, we
extract word counts from users' posts and perform PLSA topic
analysis~\cite{hofmann1999probabilistic} at the topic level. Then
given a latent topic variable $c$, each word $w$ is associated with a
probability $p(w|c)$. Second, we extract the most relevant words
from each topic, measured by the probability $p(w|c)$. Each of these
words is then annotated as a hashtag in the posts. This approach
differs from the TF-IDF \cite{chowdhury2010introduction} and other
word-frequency based methods for word importance. It enables a deeper
coverage across a range of topic areas in the posts. For example, in
addition to marking top-ranked words belonging to a large or more
popular topic as hashtags, we also include words that are highly
relevant to a small or less popular topic
cluster. Table~\ref{table:topic} shows an example list of
automatically annotated hashtags using the Twitter dataset described
in Section~\ref{section:data}.

\begin{table}[h!] 
\centering
\caption{Top-ranked words from selected topic using the topic-based method}\label{table:topic}
\smallskip\small
\vspace{-0.1in}
\begin{tabular}{|l l l | l|} 
\hline
{\bf Topic}  & {\bf Interpretation}& {\bf Automatic Hashtags} \\
\hlinewd{1pt}
1   & Jobs& job, tweetmyjobs, greater, \\
~&~&ymca, part, education, \\
~&~&boston, group    \\
\hline
2  & Food & lunch, eat, breakfast, \\
~&~&sweet, wine, cafe  \\
\hline
3  & World Cup& world, usa, top, worldcup, \\
~&~& worldcup2014, cup, goal, \\
~&~&final, beat, track \\
\hline
\end{tabular}
\end{table}

\begin{figure}[t]
\vspace{-0.05in}
\centering
\includegraphics[width=0.4\textwidth]{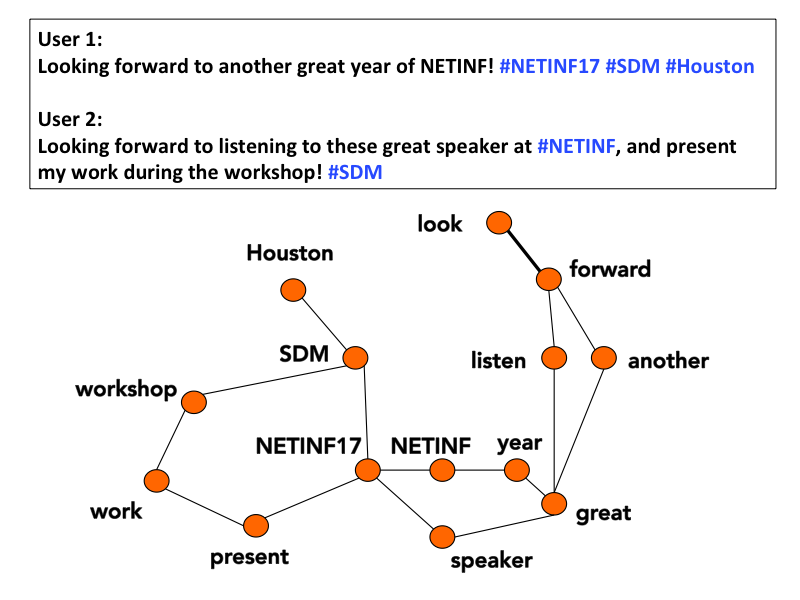}
\vspace{-0.1in}
\caption{Word co-occurrence graph construction}\label{fig:cooccurrence}
\vspace{-0.2in}
\end{figure}

Another related approach for automatic hashtag annotation is the
community-based graph method. This methods attempts to find clusters
of words via word co-occurrence relationships. First, we construct a
weighted word co-occurrence graph, where vertices in the graph
represent words and edges represent co-occurrence of words in users'
posts; see Figure~\ref{fig:cooccurrence}. We then perform community
detection on the co-occurrence graph using
infomap~\cite{lancichinetti2009community}. From each community, we
extract words with the highest PageRank values and then annotate them
as hashtags.

\subsection{Content+Context Graph Construction}\label{sec:construction}
Graph construction is performed by designating both users and the
automatically annotated hashtags as vertices and representing edges as
four types of interactions: user-to-user posts, user mentions of
hashtags, reposts and co-occurrence of users or hashtags. The count of
occurrence of different edge types is saved in the graph. For
analysis, counts are summed across edge types. The resulting content
+ context graph is weighted and undirected. It has been shown
to be useful in a number of applications. In the experiment, we construct two graphs, $G_{\rm twitter}$ for the Twitter graph and $G_{\rm inst}$ for the Instagram graph, and 
showcase its usefulness through an important application, cross-domain
entity resolution~\cite{campbell2016cross}.

\section{Experiments}
\subsection{Automatic Hashtags}
\begin{figure}[t]
\centering
\vspace{-0.08in}
\begin{subfigure}{0.8\linewidth}
\centering
\includegraphics[width=0.9\textwidth]{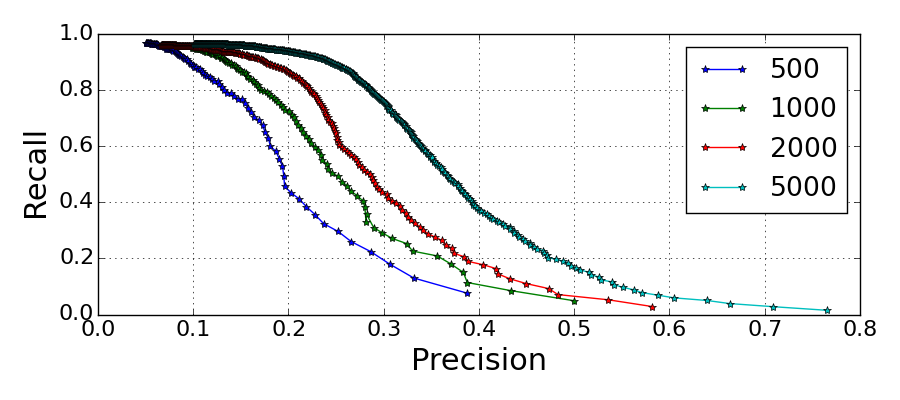}
\caption{Topic-based method}
\end{subfigure}
\begin{subfigure}{0.8\linewidth}
\centering
\includegraphics[width=0.9\textwidth]{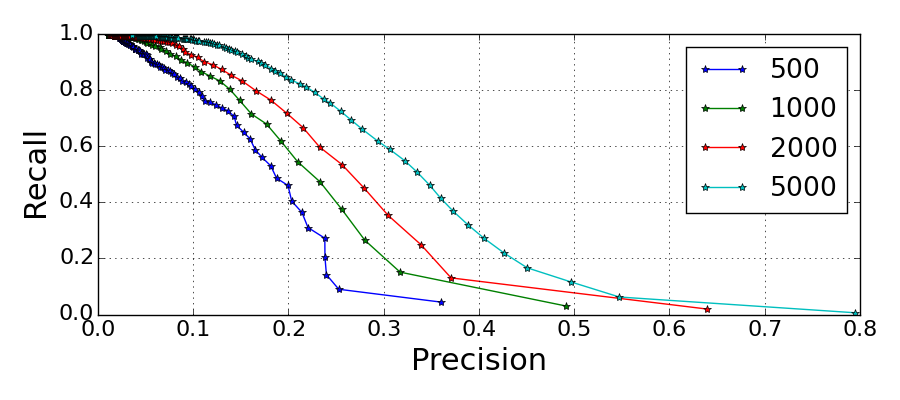}
\caption{Community-based method}
\end{subfigure}
\caption{Precision and recall curves of the automatic hashtags against top $M$ user-annotated hashtags.}\label{fig:hashtag_annotation}
\vspace{-0.2in}
\end{figure}

We perform experiments on Twitter and Instagram corpora as described
in Section \ref{section:data}. We follow the process presented in
Section~\ref{sec:preproc} for pre-processing to clean up the data and
remove user-annotated hashtags. Then, we use both the topic-based and
community-based methods for automatic hashtag
annotation. We show how the automatic hashtags compare with the user-annotated hashtags. Fig.~\ref{fig:hashtag_annotation} presents precision and
recall of recovering the hashtag vocabulary. Each curve in
Fig.~\ref{fig:hashtag_annotation} is generated for the top-$M$ most
common user-annotated hashtags where $M = 500$, $1000$, $2000$, and
$5000$.  True positives are defined as automatically selected hashtags
that are in the top-$M$ user-annotated hashtags. If $tp$ is the number
of true positives and $K$ is the number of automatically selected
hashtags, then precision is $tp/K$ and recall is $tp/M$.

The number of automatically selected hashtags, $K$, can easily be
varied with both the graph and topic methods by varying the number of
words selected per cluster.  For small $K$, precision is high, but
recall is low since less common user-annotated hashtags are missed.  For
large $K$, our methods recall all of the top-$M$ user-annotated
hashtags.  Note that for a fixed recall, as $M$ increases, precision also
increases.  This property indicates that the automatic methods are
finding user-hashtags, but not in the same order as the top-$M$ most
common user-annotated ranking.

\subsection{Cross-Domain Entity Resolution System}\label{sec:ERS}
We define cross-domain entity resolution as the problem of matching
entities across different platforms. Prior approaches to the entity
resolution problem
include~\cite{getoor:kdd13,koutra2013big,iofciu2011identifying,lyzinski2015,peled2013entity,
  campbell2016cross}.  We use the framework studied
in~\cite{campbell2016cross} to extract features for associating
entities across Twitter and Instagram.  Two classes of features are
extracted: username similarity and graph-based features. For username
similarity, we use the standard Jaro-Winkler
similarity~\cite{jaro1995probabilistic, winkler1999state} for
approximating username matches. It has been shown
in~\cite{campbell2016cross} to be best suited for username matching.

\begin{table*}
\centering\small
\caption{Summary of entity resolution results. `P' denotes the profile
  feature, `N' denotes the content + context graph feature,
  and `P\texttt{+}N' denotes the fusion of the two. The interpolated
  value is given by $^*$.}\label{table:fusion_ht}
\vspace{-0.1in}
\begin{tabular}{|l l l l l|} 
\hline
{\bf Fusion} & {\bf Hashtags}& {\bf Method} & {\bf EER ALL (\%)} & {\bf EER NT (\%)}  \\
\hlinewd{1pt}
P   &~&~&  {1.54} & {5.74}*    \\
\hline
P+N  &User-annotated Hashtags&~& {1.16}  & {3.79}   \\
\hline
P+N  &4K Automatic Hashtags&Topic& {\bf 1.17}  & {3.48}   \\
P+N  &4K Automatic Hashtags&Community& {1.19}  & {\bf 3.39}   \\
\hline
P+N  &10K Automatic Hashtags&Topic& {1.17}  & {3.51}   \\
P+N  &10K Automatic Hashtags&Community& {\bf 1.1}  & {\bf 3.24}   \\
\hline
\end{tabular}
\vspace{-0.15in}
\end{table*}

Graph-based features include community membership match and weighted
neighborhood match between pairs of Twitter and Instagram
accounts. The pipeline for computing the community membership match is
as follows. First, we use the automatically-extracted hashtags and
users' interactions to construct both a Twitter graph $G_{\rm
  twitter}$ and an Instagram graph $G_{\rm inst}$; See
Section~\ref{sec:construction}. Second, we use the technique described
in~\cite{Li2015, campbell2016cross} to merge the two graphs via common
hashtags. The resulting graph is denoted $G_{\rm merged}$; it includes
all the nodes from both $G_{\rm twitter}$ and $G_{\rm inst}$. Third,
we apply community-based detection on the merged graph $G_{\rm
  merged}$. Lastly, for each pair of users across Twitter and
Instagram, we compute their graph features: community similarity score
and weighted neighborhood similarity score. See Appendix
\ref{appendix} for more details.

The graph features are then combined with the username match to obtain
a fused entity resolution system. We use a random forest to train a
fuser with inputs being the vector of the three similarity scores for
each pair users and the output being the probability of a match. For
evaluation, we construct two sets of trials for training and testing
our system. Each trial consists of a user pairs $(U_{t}, U_{i})$ where
$U_t$ denotes a user from Twitter and $U_i$ denotes a user from
Instagram. A trial with a label `1' implies that $U_t$ and $U_i$
represent the same real-world entity, while a trial with a label `0'
means they are different entities. The performance of cross-domain
entity resolution system is measured by the standard miss and false
alarm rates. The resulting equal error rates (EER) for all trials and
non-trivial trials (NT) are shown in
Table~\ref{table:fusion_ht}. Non-trivial trials are trials where the
usernames are not an exact string match. Observe that systems using
automatic hashtag analysis perform significantly better than the
profile-only system. Also, graph construction using automatic hashtags
improves performance over a system using graphs constructed from
user-annotated hashtags.
\vspace{-0.05in}
\section{Conclusion}
We presented an end-to-end system for analyzing unstructured text data
and transforming the data into structured graphs that can be directly
applied to various downstream applications. For text analysis, we 
presented two unsupervised methods for automatic hashtag annotation: a
topic-based method and a community-based graph method. Both methods
show promising results in predicting relevant words from text. We also
build content + context graphs using the
automatically-annotated hashtags and use them for cross-domain
Twitter/Instagram entity resolution. The performance of the
resulting system outperforms the system using the original
user-annotated hashtags.
\vspace{-0.05in}

\bibliographystyle{unsrt}\scriptsize
\bibliography{ref}

\clearpage
\appendix
\normalsize
\section{Appendix}\label{appendix}
\subsection{Cross-Domain Community Detection}
We have constructed a Twitter graph
$G_{\rm twitter}$ and an Instagram graph $G_{\rm inst}$. To extract
community features, we perform cross-media community detection to
identify communities simultaneously across Twitter and Instagram
graphs. The key to achieving this is to align the graphs using {\it
seeds}. Seeds are known vertex matches across graphs (e.g., the common hashtags). We use a random
walk-based approach to align the graphs to form a single
interconnected graph. There are three general strategies: (1) {\it
aggregation} that merges pairs of vertices in the seed set; (2) {\it
linking} that adds links to the seed pairs; and (3) {\it relaxed
random walk} that allows a random walker to switch between graphs with
some probability. Once the graphs are aligned and connected, it is
straightforward to adapt Infomap~\cite{rosvall2008maps} for community
detection. Infomap is a random walk-based algorithm that partitions
the graph by minimizing an information-theoretic based objective
function.

For the experiment, we use the aggregation approach with Infomap for
community detection across Twitter and Instagram graphs. Prior
work~\cite{Li2015} shows that with a sufficient number of seeds,
the aggregation approach is the most faithful to the underlying
community structure. Specifically, we first associate a Markov
transition matrix to the {\it union} of $G_{\rm twitter}$ and $G_{\rm
inst}$. Each element in the Markov matrix represents the probability
of a random walk of length $1$ going from one vertex to the other; the
Markov transition probability is computed by normalizing the edge
weights between a vertex and all of its adjacent vertices. Second, for
each vertex pair in seeds, we merge the two vertices and update the
transition matrix with probability $p=0.5$ that a random walk moves to
the adjacent vertices in $G_{\rm twitter}$ and probability $1-p$ that
a random walk moves to the adjacent vertices in $G_{\rm inst}$. The
resulting aligned and connected graph is denoted as $G_{\rm join}$; it
includes all the vertices from both $G_{\rm twitter}$ and $G_{\rm
inst}$ and the edge weights are given by the Markov transition
matrix. Additionally, we apply Infomap only on the largest connected
component of the aligned graph $G_{\rm join}$; vertices that are in
the largest connected component have a community assignment.

\subsection{Graph-based Features}
As hinted earlier, we are interested in extracting two classes of
graph features: community features and neighborhood features. Note that
neighbors of a vertex in a graph are vertices connected by an edge to
the specified vertex; they are also referred to as 1-hop
neighbors. Generally, for a vertex $v$ in a graph, $k$-hop neighbors
are defined as vertices that are reachable from $v$ in {\it exactly}
$k$ hops.

\subsubsection{Community Features and Similarity} 
The basic idea is to be able to represent the similarity in community
membership between users across graphs. First, we perform a
cross-media community detection on Twitter and Instagram graphs. One
simple way to compare community features of two users is to assign a
value `1' to the pair that are in the same community and `0'
otherwise. However, this binary-valued similarity score will likely
cause confusion because it assigns a similarity score `1' to all users
belonging to the same community. To mitigate this problem, we propose
to represent a user's community feature via the community membership
of all its (k-hop) neighbors in its respective graph. For example, the
community feature for a Twitter user $U$ is given by a count vector
${\bf c}(U)$ with entries
\begin{equation}\label{eqn:count}
c_i = \left|\left\{N|\operatorname{comm}(N)=i,~N\in\operatorname{nbr}(U|G_{\rm twitter})\right\}\right|
\end{equation}
where $\operatorname{comm}(\cdot)$ indicates the community assignment
of vertex $N$ and $\operatorname{nbr}(\cdot)$ is the set of $k$-hop
neighbors.

For the experiment, we set $k = 1,2$ and use one of the two methods to
measure the similarity in community feature between users. One is to
compute the dot product of normalized count vectors, i.e.,
$\operatorname{sim}({\bf c}(U_i),{\bf c}(U_j))=\frac{{\bf c}(U_i)^T{\bf c}(U_j)}{\|{\bf c}(U_i)\|_2\|{\bf c}(U_j)\|_2}$. 

\subsubsection{Neighborhood Features and Similarity}
Given the aligned and connected graph $G_{\rm join}$ (i.e.,
constructed from joining $G_{\rm twitter}$ and $G_{\rm inst}$ using
seeds), we seek to compute the similarity between two users by
analyzing the proximity of their corresponding vertices in $G_{\rm
join}$. A popular approach is based on vertex
neighborhoods~\cite{Liben-Nowell:2003:LPP:956863.956972}, e.g., common-neighbors approach and its variants. The approach here is similar. However,
instead of simply counting the number of common neighbors, we
represent the neighborhood feature using the transition probability of
the random walk in $G_{\rm join}$. Specifically, the neighborhood
feature of a Twitter user $U$ is given by ${\bf p}(U)$, whose $i^{\rm
th}$ entry $p_i = p(U,U_i)$ represents the probability that a random
walk of a specified length $k$ starting at $U$ ends at the $i^{\rm
th}$ vertex $U_i$ in $G_{\rm join}$. The neighborhood similarity is
given by the normalized dot product of the neighborhood features.

We choose to use $k=1$ hop for computing the neighborhood
similarity. Note that for $k = 1$, the edge probability $p(U_1, U_2) =
0$ if $U_1$ and $U_2$ are not connected. Also, isolated vertices are
not considered.

\end{document}